\documentstyle[pra,aps]{revtex}
\def\rvec{\vec{r}}

\begin{document}
%
\title{Induced instability for boson-fermion mixed condensate of
Alkali atoms due to attractive boson-fermion interaction}
\author{T. Miyakawa, T. Suzuki and H. Yabu}
\address{Department of Physics, Tokyo Metropolitan University, 
         1-1 Minami-Ohsawa, Hachioji, Tokyo 192-0397, Japan}
\date{\today}
\maketitle
\begin{abstract}
Instabilities for boson-fermion mixed condensates 
of trapped Alkali atoms 
due to the boson-fermion attractive interaction 
are studied using a variational method. 
Three regions are shown 
for their instabilities 
according to the boson-fermion interaction strength: 
stable, meta-stable and unstable ones. 
The stability condition is obtained 
analytically from the asymptotic expansion 
of the variational total energy. 
The life-time of metastable states  
is discussed for tunneling decay, 
and is estimated to be very long. 
It suggests that, except near the unstable border, 
meta-stable mixed condensate should be almost-stable
against clusterizations. 
The critical border between meta-stable and unstable phases 
is calculated numerically 
and is shown to be consistent with the M{\o}lmer scaling condition.  
\end{abstract}
%
%
Discoveries of the Bose-Einstein condensates (BEC) 
for Alkali atoms \cite{bec,review} 
and the Fermi degeneracy for trapped ${}^{40}$K atoms \cite{DeMarco}
encourage the study for 
boson-fermion mixed condensates of trapped atoms. 
Originally, the boson-fermion condensates have been studied in 
bulk systems such as ${}^3$He-${}^4$He 
and hydrogen-deuteron systems \cite{hydro}. 
Studies of trapped mixed condensates have progressed recently;  
the static properties 
of ${}^{39,41}$K-${}^{40}$K condensate 
(with repulsive boson-boson interaction) \cite{Molmer,Nygaard,Miyakawa}, 
dynamical expansions after the removal 
of trapping potentials \cite{Amoruso}, 
and the instability changes of ${}^7$Li-${}^6$Li system 
with attractive boson-boson interaction \cite{Minniti}. 

In this paper, 
we study the instabilities and collapses 
of the polarized boson-fermion mixed condensates 
due to the attractive boson-fermion interaction. 
In BEC, such instabilities are observed in several systems:  
1) the trapped meta-stable ${}^7$Li BEC 
(observed experimentally \cite{metaLi})
equilibrated between the attractive boson-boson interaction 
and the kinetic pressure 
due to the finite confinement \cite{Baym}, 
2) two-component uniform BEC (from bosons 1 and 2),  
whose stability condition is given by 
$g_{11} g_{22} > g_{12}^2$ 
($g_{ij}$ : the coupling constant
between bosons i and j).  
In the latter case, the stability condition depends only on 
the interaction strength, 
but the condition for the case 1) has also 
the particle number dependence \cite{Baym}.

Let us consider the case of the 
trapped boson-fermion mixed condensate.  
In the present paper, 
we assume $T=0$ and 
a spherical harmonic-oscillator 
for the trapping potential; 
extension to the deformed potential is straightforward. 
Using the Thomas-Fermi approximation 
for the fermion degree of freedom 
(appropriate for large $N_f$ condensate), 
the total energy of the system 
becomes a functional of  
the boson order-parameter $\Phi(\vec{r})$ 
and the fermion density distribution $n_f(\vec{r})$:  
\begin{eqnarray}
  E[\Phi(\vec{r}),n_f(\vec{r})] =\int{\,d^3r} 
    \left[ \frac{\hbar^2}{2m_b} | \nabla\Phi(\vec{r})|^2
          +\frac{1}{2} m \omega^2 \vec{r}^2 |\Phi(\rvec)|^2 
          +\frac{g}{2} |\Phi(\vec{r})|^4 
    \right] \nonumber \\ 
  +\int{\,d^3r} 
    \left[ \frac{3}{5} \frac{\hbar^2}{2m} 
           (6\pi^2)^{\frac{2}{3}} 
           n^{\frac{5}{3}}_f(\vec{r})
          +\frac{1}{2} m \omega^2 \vec{r}^2 n_f(\vec{r}) 
          +h |\Phi(\vec{r})|^2 n_f(\vec{r}) 
    \right],
\label{eQc}
\end{eqnarray}
where $\omega$ and $m$ are 
the angular frequency for the boson/fermion trapping potential 
and the boson/fermion mass 
($\omega_b =\omega_f \equiv \omega$ and  
and $m_b =m_f \equiv m$ 
are assumed for simplicity). 
In low-density gas, 
the boson-boson/boson-fermion coupling constants 
$g$, $h$ are represented 
by the s-wave scattering lengths $a_{bb}$ and $a_{bf}$: 
$g =4\pi \hbar^2 a_{bb}/m_b$ and 
$h =2\pi \hbar^2 a_{bf}/m_r$ 
where $m_r =m_b m_f/(m_b+m_f)$. 
In eq. (\ref{eQc}), 
the fermion-fermion interaction has been neglected;
the elastic fermion-fermion s-wave scattering for a polarized gas
is absent because of the Pauli blocking effects 
and the p-wave scattering is suppressed 
below 100$\,\mu{\rm K}$ \cite{DeMarco2}. 

Evaluating the total energy 
with the variational method,  
we take the Gaussian ansatz 
for the boson order-parameter: 
\begin{equation}
  \Phi(x;R) \equiv \xi^{-3/2} \phi(x;R) 
            =\sqrt{\left(\frac{3}{2\pi}\right)^{\frac{3}{2}} 
                   \frac{N_b}{(\xi R)^3} }
               \exp\left(\frac{-3 x^2}{4 R^2} \right), 
\label{eQd} 
\end{equation}
where $x =r/\xi$ is a radial distance scaled 
by the harmonic-oscillator length $\xi =(\hbar/m \omega)^{1/2}$, 
and $N_b =\int{d^3x} |\Phi(x;R)|^2$ is total boson number. 
The variational parameter $R$ in (\ref{eQd})  
corresponds to the root-mean-square (rms) radius 
of the boson density distribution. 
This kind of variational functions have also been used 
in the instability investigations of ${}^7$Li-BEC \cite{Baym} 
and ${}^7$Li-${}^6$Li system \cite{Minniti}. 
In the Thomas-Fermi approximation 
with (\ref{eQd}), 
the fermion density distribution $n_f(\vec{r})$ becomes
\begin{equation}
  n_f(x;R) =\frac{1}{6\pi^2}
            \left[ {\tilde \mu}_f(R) 
                  - x^2 
                  -{\tilde h} |\phi(x;R)|^2  
            \right]^{\frac{3}{2}}. 
\label{eQe}
\end{equation}
where 
${\tilde h} =2h/(\hbar \omega \xi^3)$.  
This fermion density vanishes at the turning point
$\Lambda$ which is determined by 
\begin{equation}
  F(\Lambda) \equiv 
   {\tilde \mu}_f -\Lambda^2 
   -\frac{C}{R^3} e^{-\frac{3 \Lambda^2}{2 R^2}} =0,   
\label{eQf}
\end{equation}
where $C =(3/2\pi)^{3/2} {\tilde h} N_b$. 
We take $n_f(x;R) \equiv 0$ when $x \geq \Lambda$. 

The scaled chemical potential 
${\tilde \mu}_f =2 \mu_f/(\hbar \omega)$ 
in eq. (\ref{eQe}) is determined by the normalization condition 
$\int{d^3x} n_f(x) =N_f$. 
As a result, the two parameters $\Lambda$ and ${\tilde \mu}_f$ 
become functions of $R$. 

Using eqs. (\ref{eQd}) and (\ref{eQe}), 
the total energy $E(R) =E[\Phi(x;R),n_f(x;R)]$ becomes 
\begin{eqnarray}
  E(R)/(\hbar \omega) 
    &=& N_b \left\{ \frac{9}{8} \frac{1}{R^2} 
                  +\frac{1}{2} R^2
                  +2^{-5/2} 
                   \frac{3}{\pi} 
                    {\tilde g} N_b \frac{1}{R^3} \right\}  
     \nonumber\\ 
    &+& N_f \left\{ I_{ke}(R) + I_{ho}(R) 
                  +C R^{-3} I_{bf}(R) \right\}, 
\label{eQg}
\end{eqnarray}
where ${\tilde g} =2g/(\hbar \omega \xi^3)$ 
and integrals $I_{ke,ho,bf}(R)$ are defined by 
\begin{eqnarray}
  I_{ke}(R) &=& \frac{1}{5 \pi N_f} 
              \int_0^\Lambda dx x^2 F(x)^{5/2},  \nonumber\\
  I_{ho}(R) &=& \frac{1}{3 \pi N_f} 
              \int_0^\Lambda dx x^4 F(x)^{3/2},  \nonumber\\
  I_{bf}(R) &=& \frac{1}{3 \pi N_f}
              \int_0^\Lambda dx x^2 e^{-\frac{3 x^2}{2 R^2}} F(x)^{3/2}. 
\label{eQh}
\end{eqnarray}

One of the most important candidates for mixed condensate 
is the ${}^{39,41}$K-${}^{40}$K (boson-fermion) system. 
Their scattering lengths are not well fixed at present, 
and different values have been reported experimentally \cite{Cote,Kscat1,Kscat2}.  
In the present calculations, 
to estimate the qualitative stability behaviors,  
we take the value for $a_{bb}({}^{41}{\rm K})$ in ref.\cite{Cote}, 
and it gives ${\tilde g} =0.2$ for the boson-boson interactions 
with $\omega =450{\rm\,Hz}$. 
For the attractive boson-fermion interaction ${\tilde h}$, 
several negative values are taken: 
$\alpha \equiv |{\tilde h}|/{\tilde g} =(0 \sim 3)$.  
It should be noted that the interaction strength 
can be shifted using the Feshbach resonance phenomena \cite{feshbach},  
whose applications for K-atom have been discussed in \cite{Kscat1,Kfesh}. 

In fig.~1, numerical results for $E(R)$ in (\ref{eQg}) 
are plotted for
$\alpha =|{\tilde h}|/{\tilde g} =0.1$, $1.0$, 
$2.0$, $2.5$, $3$ (lines a-e) with $N =10^6$, 
where three kinds of patterns can be read off:  
stable (a), meta-stable (b,c) and unstable states (d,e).  
In weak boson-fermion interaction, the system is stable 
against the variation of $R$, 
and has an absolute minimum as an equilibrium state. 
For the intermediate strength, 
it becomes meta-stable with one local minimum (lines b,c), 
and, in strong attractive interactions, 
the minimum disappears and the system become unstable (lines d,e). 
It should be noted that similar meta-stable states appear  
in the BEC with attractive interactions (${}^7$Li) \cite{Baym},   
but no stable states can exist in that case. 

As can be found in fig.~1, 
the stability of mixed condensate can be judged 
from the small $R$ behavior of $E(R)$: 
positive divergence in small $R$ suggests
stable state. 
To obtain the stability condition, 
we consider analytically the asymptotic expansion of $E(R)$ 
in $R \ll 1$. 
For that purpose, 
the leading order terms of ${\tilde \mu}_f(R)$ and $\Lambda(R)$
in small $R$ 
should be determined from eq. (\ref{eQf}) 
and the normalization condition: 
$\int{d^3x} n_f(x;R) =N_f$. 
To evaluate them, 
we assume $\Lambda(R)/R \ll 1$ when $R \ll 1$, 
and expand the Gaussian function 
to the order of $(x/R)^2$: 
$\exp(-3 x^2/2 R^2) \sim 1 -(3 x^2/2 R^2)$. 
The consistency of these assumptions will be shown later, 
and can be also checked numerically. 
These assumptions makes  
the integral in the normalization condition 
evaluated analytically, and we obtain 
\begin{eqnarray}
\label{mu0}
  \tilde{\mu} &\sim& -\left( \frac{3}{2\pi} \right)^{\frac{3}{2}} 
                      |{\tilde h}|N_b \frac{1}{R^3} 
                     +\sqrt{\frac{3}{2}}
                      \left( \frac{3}{2\pi} \right)^{\frac{3}{4}} 
                     (48 N_f)^{\frac{1}{3}} 
                     ({\tilde h} N_b)^{\frac{1}{2}} 
                     \frac{1}{R^{\frac{5}{2}}}, 
\label{eQi}\\
  \Lambda    &\sim& \left( \frac{2}{3} \right)^{\frac{1}{4}} 
                    \left( \frac{2\pi}{3} \right)^{\frac{3}{8}} 
                    \frac{(48 N_f)^{\frac{1}{6}}}{
                          ({\tilde h} N_b)^{\frac{1}{4}}} 
                    R^{\frac{5}{4}}. 
\label{eQj}
\end{eqnarray}
It should be noted that 
eq. (\ref{eQj}) leads to 
$\Lambda/R \propto R^{1/4} \ll 1$ for $R \ll 1$, 
which shows the consistency of our assumption. 
Using eqs. (\ref{eQi}) and (\ref{eQj}), 
in the leading order in small $R$, 
the $E(R)$ becomes
\begin{eqnarray}
  E(R) /(\hbar\omega) 
    &\sim& \left( \frac{3}{2\pi} \right)^{\frac{3}{2}}
           \frac{N_b}{2} 
           \left\{ \frac{{\tilde g} N_b}{2^{\frac{2}{5}}} 
          -|{\tilde h}| N_f \right\} \frac{1}{R^3} \nonumber\\
        &+&\frac{1}{4} \frac{9}{6^{\frac{1}{6}}} 
         \left( \frac{3}{2\pi} \right)^{\frac{3}{4}} 
         (|{\tilde h}| N_b)^{\frac{1}{2}} 
         N_f^{\frac{4}{3}} \frac{1}{R^{\frac{5}{2}}} 
        +{\cal O}(R^{-2}), 
\label{eQk}
\end{eqnarray}
so that, in small $R$ region, 
$E(R)$ is dominated by $R^{-3}$-term, 
and its positivity gives the stability condition: 
\begin{equation}
  \alpha \equiv \frac{|{\tilde h}|}{{\tilde g}}
         < 2^{-5/2} \frac{N_b}{N_f}. 
\label{eQkAAA}
\end{equation} 
For the cases of fig.~1 , it becomes $\alpha < 0.18$, 
which shows that only line (a) is for stable state. 
This condition is also plotted as the dashed line in fig.2 .

From the above stability condition, 
lines (b) and (c) in fig.~1 are found to be meta-stable, 
so that, in principle, 
the equilibrium states (at local minima $R=R_{eq}$) 
should collapse by the quantum tunneling effects 
into those of $R=0$ 
through the potential barrier 
between $R=0$ and $R_{eq}$.  
To study the collapses of meta-stable states, 
we estimate their collective tunneling life-time 
$\tau_{ct}$ applying the Gamow-theory of the nuclear $\alpha$-decay 
to these states. 
For the order estimation of the tunneling life-time, 
as an approximation for $E(R)$, 
we use the simplified linear-plus-harmonic oscillator potential 
$V(R)$: 
\begin{equation}
\label{eQkAAAA}
V(R) = \cases{V_1(R) =F R -G,  
                  &($0 \le R < R_t$)  \cr
                  V_2(R) =\frac{1}{2} 
                          M \Omega^2 (R-R_{eq})^2 
                         +V_0.  
                  &($R_{t} \le R$)    \cr}
\end{equation}
As shown in appendix A, 
the tunneling life time $\tau_{ct}$ for $V(R)$ becomes 
$\tau^{-1}_{ct} =D \exp(-W)$,   
where $D$ is a staying probability on barrier surface per unit time 
and $\exp(-W)$ is a transmission coefficient for the potential barrier. 
The explicit formulae for $D$ and $\exp(-W)$ are 
\begin{eqnarray}
    D &=&\frac{4\sqrt{2}\Omega}{\sqrt{\pi}} 
         \sqrt{ \frac{\Delta V}{\hbar\Omega} 
               -\frac{1}{2}} 
\label{eQkAAB}\\
    W &=&\frac{4}{3} 
         \sqrt{\frac{2M}{\hbar^{2}}} 
         \left( \Delta{V} -\frac{\hbar\Omega}{2} \right) 
         (R_t -R_E) 
        +\frac{(R_0 -R_t)^2}{a_{HO}^{2}}, 
\label{eQkAAC}
\end{eqnarray}
where $R_E$ is a WKB turning point for the energy $E$ 
and $\Delta{V} = V_1(R_t) - V_0$ is a barrier height  
(see fig.~3). $a_{HO}=\sqrt{\hbar/M\Omega}$ is harmonic
oscillator length around equilibrium point.
The derivation of eqs. (\ref{eQkAAB},\ref{eQkAAC}) 
is given in appendix A. 

To apply the above formula for the metastable condensates, 
we take $M =N m$ and $\Omega =2 \omega$. 
These values have been determined to be the values in
harmonic oscillator potential around the equilibrium point
in the non-interacting case.
For parameters ($R_0$,$V_0$), ($R_t$,$V_1(R_t)$) and $R_E$, 
we use the values that can be obtained from 
the numerically calculated $E(R)$: 
the equilibrium, maximum, WKB turning points 
(for $E =V_0 +\hbar\Omega/2$). 

For the present case of $N =10^6$ and $\alpha =2.0$
we obtain $R_t -R_E \sim 1$, $R_0 -R_t \sim 3$ 
in the unit of $\hbar/m\omega$, 
and $\Delta{V} \sim 40$ in the unit of $N\hbar\omega$ 
(see the line (c) in fig.~1). 
Using these values for eqs. (\ref{eQkAAB},\ref{eQkAAC}), 
the staying probability and the transmission coefficient 
become $D \sim 10^8 \omega$ and $W \sim 10^7$. 
As a result, we obtain 
$\tau^{-1}_{ct}/\omega \sim 10^8 \times \exp(-10^7)$, 
and the life-time $\tau_{ct}$ becomes very longer 
than, for example, 
that of clusterization by many-body collisions, 
$\sim (1 \sim 10){\rm\,sec}$. 
Thus, meta-stable states should be really 
almost stable
except the ones extremely close to the unstable border 
(such as line d). 
This long lifetime has originated from collectiveness 
of very many particles, 
and is consistent with that for meta-stable BEC 
\cite{Tunneling}. 
To shorten the tunneling life-time and 
make meta-stable states collapse, 
the parametric excitation with $\omega =\omega(t)$ may
give an interesting possibility,  
which will be discussed in another paper \cite{MiyaN}. 

For unstable states (e.~g. line e in fig.~1), 
using eqs. (\ref{eQi},\ref{eQj}), 
the leading $R$-dependence 
of the fermion density at $x=0$ and its rms radius are calculated:  
$n_f(0) \propto R^{-\frac{15}{4}}$ 
and $(x_f)_{rms} \propto R^{\frac{5}{4}}$,  
which are also checked numerically. 
They show that,  
in the collapses of metastable and unstable states, 
both boson and fermion distributions become localized 
and compressed, just like gravitational collapses 
of massive stars. 
Those collapses can also be found in  
the instability by the collective excitations 
evaluated with the sum-rule method \cite{Miya}, 
where the boson-fermion in-phase monopole excitation   
becomes the zero mode. 

Finally, we comment about the critical condition 
for the unstable regions, 
which has originally been obtained by M{\o}lmer \cite{Molmer} 
with the Thomas-Fermi approximations for both fermion and boson 
distribution in form of the scaling relation: 
$h^2/g \propto N_f^{-p/(3p+6)}$ 
for $V \propto r^p$-type trapping potential
(the derivation of the scaling law by M{\o}lmer, 
see appendix B). 
For the harmonic oscillator potential ($p=2$), 
it becomes $\alpha^2 N_f^{1/6} ={\rm const}/g$.  
In the present framework of the variational method, 
the critical $\alpha$ between the meta-stable 
and unstable states have been evaluated for several $N$ 
and plotted in fig.~2 (filled circles). 
From them, 
we can find that the variational results 
are also consistent with the M{\o}lmer scaling law: 
the solid line added in fig.~2 in the unstable region.  

In summary, 
for stability of boson-fermion 
mixed condensates with boson-boson repulsive and 
boson-fermion attractive interactions, 
the present variational calculations give
the phase diagram with three regions: 
stable, meta-stable and unstable ones. 
They are clearly shown in fig.~2 
in the case of $N_f =N_b \equiv N$ and 
$\omega_f =\omega_b \equiv \omega$.  
To estimate borders of stability regions generally, 
we studied their critical conditions analytically,   
from which we can find that 
the diagram should have qualitatively the similar pattern 
also in general cases of the mixed condensates. 
Finally, it should be noted that, 
in the boson-fermion collapsing processes 
of unstable and meta-stable states, 
the increasing boson/fermion densities 
make the two-body short-range interactions 
(e.g. $p$-wave scattering processes) 
and three-body (or higher) collisions 
more effective. 
These interactions, 
which are not essential for low-density condensates, 
may play important roles 
through the fermion-paired superfluid formations 
and clusterization into metal states 
by tunneling phenomena \cite{Tunneling}. 
They will be discussed in a future publication \cite{MiyaN}. 
%
%
\appendix
\section{Life Time of Collective Tunneling Effect in Metastable Region}
To estimate a life time of metastable mixed condensates, 
we consider the collective tunneling effect 
with the collective variable $R$ (the boson radius) 
and its effective energy $E(R)$ as a collective potential. 
The effective energy $E(R)$ is obtained from (\ref{eQk}), 
and, for the metastable condensates, 
a typical shape of it is the line (c) in Fig.~1. 
We approximate this $E(R)$ 
with the linear-plus-harmonic-oscillator type potential eq. 
(\ref{eQkAAAA}) (Fig.~3).
%
$M$ is inertia mass for the collective variable $R$. 
The meanings of other parameters 
($F$, $G$, $\Omega$, $R_{eq}$, $R_t$) 
can be read off in Fig.~3, 
and they are fixed in order that the potential $V(R)$ reproduce 
the $E(R)$ numerically obtained by (\ref{eQk}). 
This approximation should be enough for the order estimation 
of the life time. 
For the kinetic term, we take
\begin{equation}
     T =-\frac{\hbar^2}{2M} \frac{d^2}{dR^2}. 
\label{aP2}
\end{equation}

In $V(R)$, the metastable state $\psi_M(x)$ before tunneling 
is approximately given by the ground-state wave function 
in the harmonic oscillator potential $V_2(R)$:
\begin{equation}
     \psi_M(R) = \frac{1}{\sqrt{\pi^{\frac{1}{2}} a_{HO}}}
     \exp{\left[-\frac{(R-R_{eq})^2}{a_{HO}^2} \right]}, 
\label{aP3}
\end{equation}
where $a_{HO}=\sqrt{\hbar/M\Omega}$ is a harmonic oscillator length. 
This state has the zero-point energy $\hbar\Omega/2$
measured from $V_{0}$.
The continuum state $\psi_D(R)$ after the tunneling decay is obtained by 
the wave function in the linear potential $V_1(R)$, 
and its Schr\"odinger equation becomes 
\begin{equation}
     \left[ \frac{\hbar^2}{2M} \frac{d^{2}}{dR^2} 
           +E -(F R -G) \right] \psi_D(R) =0. 
\label{aP4}
\end{equation}
To solve the above equation, 
we use the WKB approximation. 
It should be noticed that, 
because the state energy is lower than the potential maximum 
($E < F R_t -G$), 
a turning point exist at $R_E =(E +G)/F$. 
Thus, the WKB connection formula should be used at $R_E$ 
for the continuum state $\psi_D$: 
\begin{eqnarray}
     \psi_D(R) &=&\frac{A}{2 \sqrt{K(R)}}
          \sin{\left[ \frac{2}{3} 
                      \left( \frac{2MF}{\hbar^2} \right)^{\frac{1}{2}}
	              (R_E -R)^{\frac{3}{2}} 
                     +\frac{\pi}{4} \right]}, 
          \quad \hbox{for } 0 \le R < R_E 
\label{aP5}\\
     \psi_D(R) &=&\frac{A}{2 \sqrt{K(R)}}
          \exp{\left[-\frac{2}{3} 
                      \left( \frac{2MF}{\hbar^2} \right)^{\frac{1}{2}}
                      (R-R_{E})^{\frac{3}{2}} \right]}, 
          \quad R_E < R
\label{aP6}
\end{eqnarray}
where
$A =\sqrt{ 2M/(\pi\hbar^2) }$ and 
$\hbar K(R) =\sqrt{2MF |R_E -R|}$. 

Let us consider the tunneling decay rate 
$\Gamma$ from $\psi_M(R)$ to $\psi_D(R)$. 
Using the golden rule of Fermi, 
it becomes
\begin{equation}
     \Gamma =\frac{2\pi}{\hbar} 
             \left| \int dR \, 
                     \psi_D^*(R) [V(R) -V_2(R)] 
                     \psi_M(R) \right|^2_{E =E_M}, 
\label{aP7}
\end{equation}
where $E_M =V_0 +\hbar \Omega/2$ is the energy of $\psi_M(R)$. 
Because $V(R) -V_2(R) =0$ for $R_t < R$ and $\psi_M(0),
\psi'_M(0) \ll 1$, eq. (\ref{aP7}) becomes 
\begin{equation}
    \Gamma =\frac{2\pi}{\hbar} 
              \left( \frac{\hbar^2}{2M} \right)^2
              \left[ \frac{d \psi_D(R)}{dR} \psi_M(R) 
                    -\psi_D(R) \frac{d \psi_M(R)}{dR} 
              \right]_{R=R_t}. 
\label{aP8}
\end{equation}
Using eqs. (\ref{aP3}, \ref{aP5}, \ref{aP6}), 
the tunneling rate $\Gamma$ is shown to have the form 
$\Gamma =D \exp(-W)$ : $D$ can be interpreted as 
the staying probability Per unit time 
and $\exp(-W)$ is the transition coefficient. 
The explicit formulae for $D$ and $W$ are
\begin{eqnarray}
    D &=&\frac{4\sqrt{2}\Omega}{\sqrt{\pi}} 
         \sqrt{ \frac{\Delta V}{\hbar\Omega} 
               -\frac{1}{2}} 
\label{aP9}\\
    W &=&\frac{4}{3} 
         \sqrt{\frac{2M}{\hbar^{2}}} 
         \left( \Delta{V} -\frac{\hbar\Omega}{2} \right) 
         (R_t -R_E) 
        +\frac{(R_0 -R_t)^2}{a_{HO}^{2}}, 
\label{aP10}
\end{eqnarray}
where $\Delta{V} = V_1(R_t) - V_0$ is a barrier height.  
To represent the parameters $F$ and $G$ in $V_1$ 
by $\Delta{V}$, 
we have used 
\begin{equation}
     V_1(R_t) =F R_t -G,  \quad
     V_1(R_E) =F R_E -G =E =V_0 +\frac{1}{2} \hbar \Omega. 
\label{aP10A}
\end{equation}
The tunneling life-time is defined as an inverse of decay rate: 
$\tau_{ct} = \Gamma^{-1}$.
\section{Critical Condition for Unstable Region}
In this appendix, we rederive the M{\o}lmer scaling law 
which gives the critical condition for the unstable region \cite{Molmer}. 
We assume the Thomas-Fermi approximations both 
for the boson and fermion density distributions. 
In that approximations, the density distributions are given as 
solutions of the Thomas-Fermi equations: 
\begin{eqnarray}
  & {\tilde g} n_b(x) +x^2 +{\tilde h} n_f(x) ={\tilde \mu}_b, 
\label{ap11}\\
  & [6\pi^2 n_f(x)]^{{2 \over 3}} +x^2 
      +{\tilde h} n_b(x) ={\tilde \mu}_f, 
\label{ap12}
\end{eqnarray}
where $n_b(x)$ is the boson density distribution 
scaled by the harmonic-oscillator length $\xi =(\hbar/m\omega)^{1/2}$ 
($n_b(x) =|\Phi(x)|^2 \xi^3$). 
The scaled chemical potentials, ${\tilde \mu}_{b,f}$,  
and coupling constants, ${\tilde g}$ and ${\tilde h}$, 
have been defined in the main body of this paper.

Eliminating $n_b(x)$ in (\ref{ap12}) by (\ref{ap11}), 
we obtain the equation $F[n_f(x)] =G[n_f(x)]$ 
to determine the fermion density $n_f(x)$, 
where  
\begin{equation}
     F[n_f(x)] \equiv [6\pi^2 n_f(x)]^{{2 \over 3}}, \quad
     G[n_f(x)] \equiv {\tilde \mu}_f 
       +{|{\tilde h}| \over {\tilde g}} {\tilde \mu}_b 
       -\left(1+{|{\tilde h}| \over {\tilde g}}\right) x^2 
       +{|{\tilde h}|^2 \over {\tilde g}} n_f(x). 
\label{ap13}
\end{equation}
We concentrate on the central density $n_f(0)$. 
In order to determine $n_{f}(0)$ two conditions
should be satisfied: 
\begin{equation}
     F[n_f(0)] =G[n_f(0)],    \quad\quad 
     {\delta F \over \delta n_f}[n_f(0)] 
          \ge {\delta G \over \delta n_f}[n_f(0)],  
\label{ap14}
\end{equation}
Evaluating equations in (\ref{ap14}) with (\ref{ap13}), 
we obtain the critical condition  for the unstable region:
\begin{equation}
     {|{\tilde h}|^2 \over {\tilde g}} 
          ={4\pi^2 \over \sqrt{3}} 
           \left( {\tilde \mu}_f 
                 +{|{\tilde h}| \over {\tilde g}} 
                  {\tilde \mu}_b \right). 
\label{ap15}
\end{equation}
To evaluate the right-hand side of (\ref{ap15}), 
we should use the relations between ${\tilde \mu}_{b,f}$ 
and the boson/fermion particle number $N_{b,f}$, 
which are obtained by solving the Thomas-Fermi equations in 
(\ref{ap12}). 
Here we assume that ${\tilde \mu}_b =0$ and 
${\tilde \mu}_f =2 (6 N_f)^{1/3}$ 
(${\tilde \mu}_f$ for a free fermion system) 
for (\ref{ap15}). 
Consequently, we obtain the M{\o}lmer scaling relation: 
\begin{equation}
     {|{\tilde h}|^2 \over {\tilde g}} 
          ={4\pi^2 \over 6 ^{1/6} \sqrt{6}} N_f^{-1/2} 
          \sim 12.0 N_f^{-1/6}. 
\label{ap16}
\end{equation}
It should be noted that the coefficient 12.0 in (\ref{ap16}) 
are close to the value 13.8 
which is obtained by M{\o}lmer \cite{Molmer} with numerically evaluating 
the Thomas-Fermi equations. 
%
%

\newpage
%
%
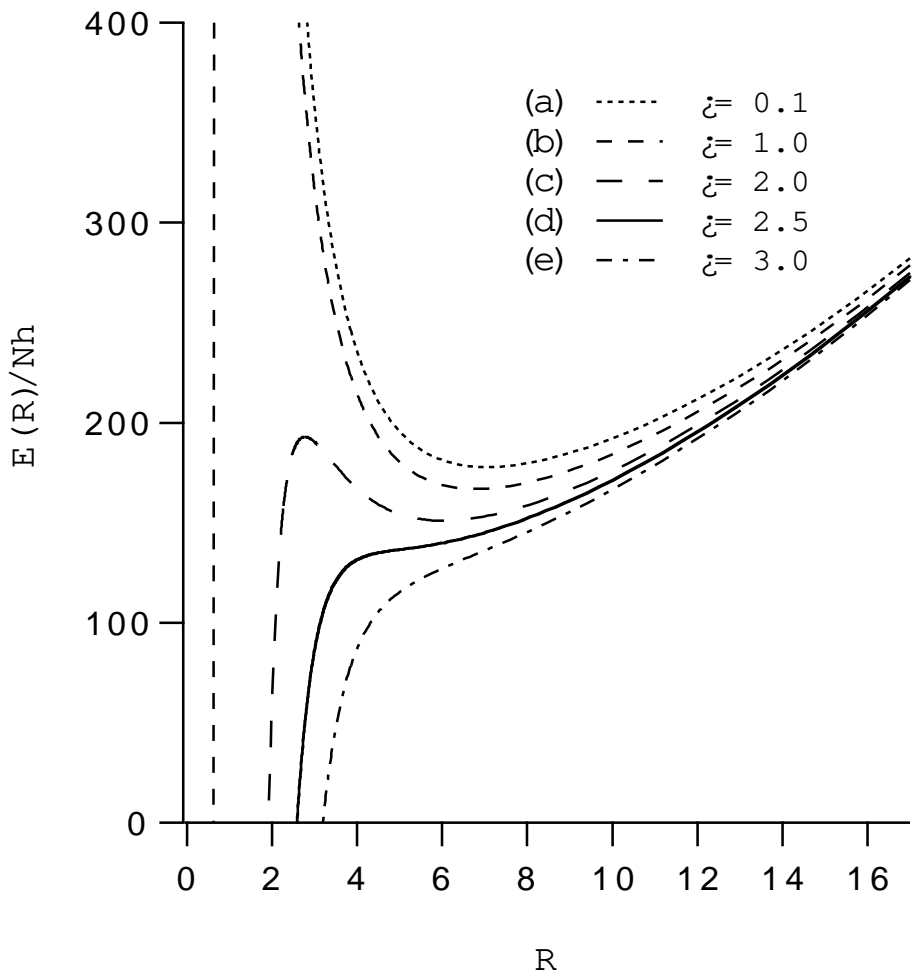
\begin{figure}
\caption{The total energy variation $E(R)/(N\hbar\omega)$ 
of the boson-fermion mixed condensate 
with the Gaussian boson distribution $\Phi(x;R)$ of radius $R$: 
$E(R)$ and $\Phi(x;R)$ are defined in (\ref{eQg}, \ref{eQd}). 
The Thomas-Fermi density function (\ref{eQe}) is applied for 
the fermion distribution. 
$N =N_b=N_f =10^6$, 
${\tilde g} =0.2$, 
and 
$\alpha ={\tilde h}/{\tilde g} =0.1$, $1.0$, $2$, $2.5$, $3$ 
for lines a-e.} 
\end{figure}

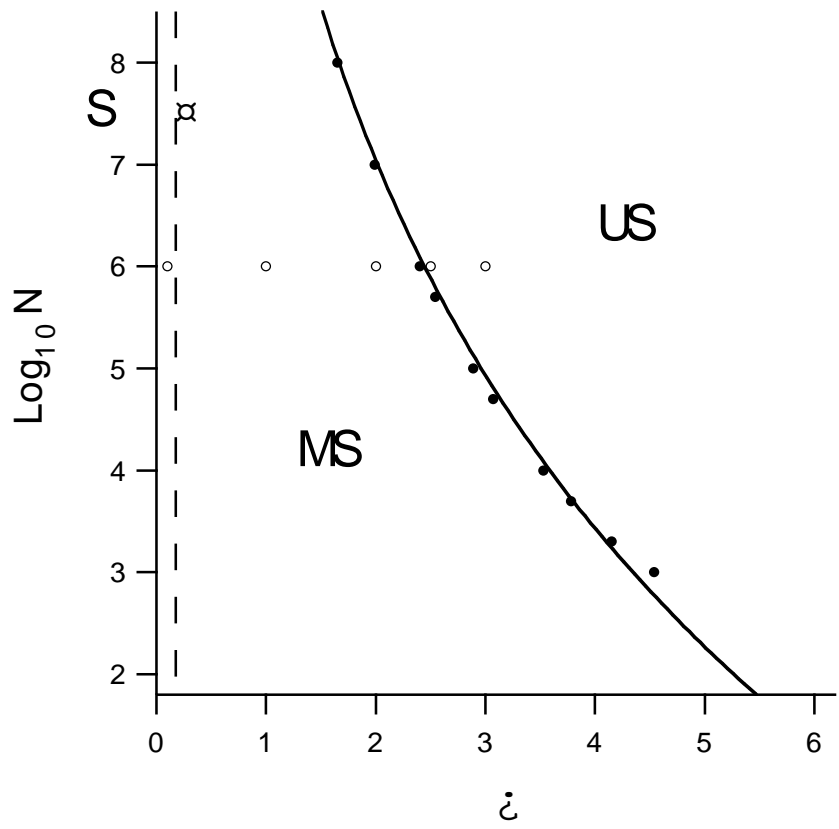
\begin{figure}
\caption{Stability phase diagram 
of the boson-fermion mixed condensate 
for $N \equiv N_b =N_f$ 
and $\alpha ={\tilde h}/{\tilde g}$  
when ${\tilde g} =0.2$.  
Three regions exist in it:  
stable (S), meta-stable (MS) and unstable (US) ones. 
The dashed line at $\alpha =0.18$ corresponds to 
the border of stable region, 
and the solid line shows the M{\o}lmer scaling law  
between meta-stable and unstable regions. 
The open circles correspond to the parameters 
of the states in fig.~1, 
and the filled circles show 
numerically confirmed critical states 
between meta-stable and unstable regions. }
\end{figure}
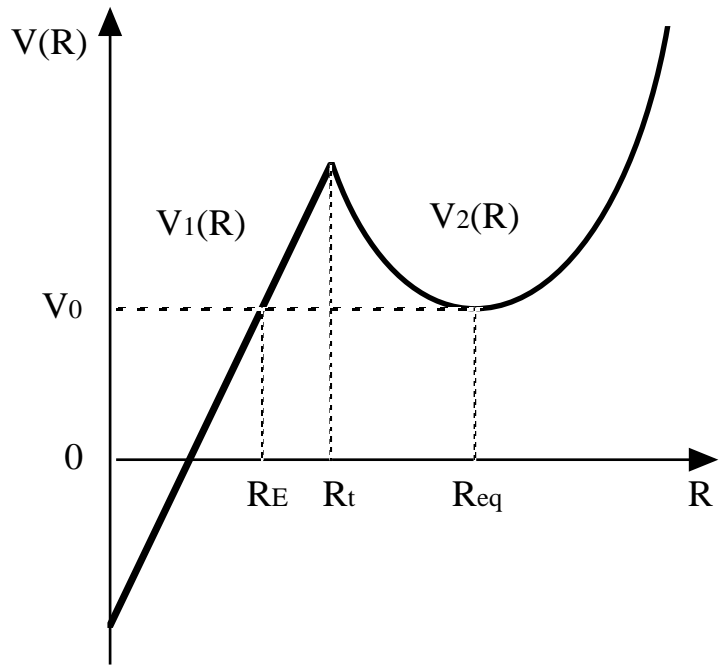
\begin{figure}
\caption{Simplified potential for meta-stable condensates.
The $V_{1}(R)$ and $V_{2}(R)$ are the linear and harmonic oscillator 
parts of the potential $V(R)$. 
The $R_{eq}$, $R_t$, $R_E$ are the equilibrium, maximum 
and the WKB turning points. 
$V_0$ is the equilibrium energy of the potential.}
\end{figure}
\end{document}